\documentclass[a4paper,10pt,twoside]{article}
\usepackage{amsmath} %I used this for a couple of nonstandard symbols.
\usepackage{vmargin} %Defaults margins
\usepackage{epsfig}%Postscript figure inclusion.
\usepackage{subfigure} %Labels subfigures within a figure environment.
\begin{document}
%========================Define new commands. 
%Some of these may be useful to you.
%I found some to be invaluable. This is all standard LaTeX.
\newcommand{\um}{\ensuremath{~\mu\mathrm{m}}}
\newcommand{\pcm}{cm\ensuremath{^{-1}}}
\newcommand{\ee}[2]{\ensuremath{#1\times10^{#2}}}
\newcommand{\degC}{\ensuremath{^\circ}C}
\newcommand{\degrees}{\ensuremath{^\circ}}
\newcommand{\tilith}{Ti:LiNbO\ensuremath{_3}}
\newcommand{\Hplus}{H\ensuremath{^+}}
\newcommand{\HHplus}{H\ensuremath{_2^+}}
\newcommand{\Heplus}{He\ensuremath{^+}}
\newcommand{\myvector}[1]{\ensuremath{\mathbf{#1}}}
\newcommand{\myemph}[1]{\textit{#1}}
\newcommand{\etal}{\textit{et~al.}}
\newcommand{\pslash}{/\!\!\!p}
\newcommand{\qslash}{/\!\!\!q}
\newcommand{\kslash}{/\!\!\!k}
\newcommand{\beqa}{\begin{eqnarray}}
\newcommand{\eeqa}{\end{eqnarray}}
\newcommand{\flux}{{\rm cm^{-2}s^{-1}sr^{-1}}}
\newcommand{\tabnum}[1]{\ref{1}}
\newcommand{\bea}{\begin{eqnarray}}
\newcommand{\eea}{\end{eqnarray}}
\newcommand{\nn}{\nonumber}
\newcommand{\V}{\vert}
\newcommand{\db}{\bar{d}}
\newcommand{\eb}{\bar{e}}
\newcommand{\nb}{\bar{\nu}}
\newcommand{\ub}{\bar{u}}
\newcommand{\vb}{\bar{v}}
\newcommand{\delslash}{/\!\!\!\partial}
\newcommand{\Eq}{Eq.~}
\newcommand{\Fig}{Fig.~}
\newcommand{\Ham}{Hamiltionian\ }
\newcommand{\Lor}{Lorrentz\ }
\newcommand{\chap}{Chap.}
\newcommand{\En}{E_{\nu_{l}i}}
\newcommand{\ezero}{{\cal E}_0}
\title{Neutrino production, oscillation and detection in the presence of general four-fermion interactions} 
\author{Matthew Garbutt \& Bruce H.~J.~McKellar\\
School of Physics, University of Melbourne, Victoria, Australia 3010}
\date{28 July 2003}
\maketitle
\begin{abstract}
In this note we investigate the impact of new physics in the form of
general four-fermion interactions on neutrino oscillation signals.  We
develop a field theoretic description of the overall oscillation
process which includes the non-standard interactions during the
neutrino production, propagation and detection stages.  Insights
gained  during the development of this formalism 
 regarding the possibility of new interactions mimicking neutrino
mass differences  are
expounded.   The impact of
possible new physics is assessed by studying the $\nu_\mu\to\nu_\tau$
oscillation channel in vacuum and the $\nu_e\to \nu_\mu$ channel in
matter. 
Although it is known that the effects of new interactions can only
act as a perturbation to the leading oscillation parameters, we find
that  great
care needs to be taken when drawing conclusions regarding the strength
of new couplings from oscillation measurements.  
\end{abstract}
\section{Introduction}

The observation of neutrino oscillations have had a profound impact on
our understanding of particle physics.  The 
measurement of atmospheric neutrino oscillations at the
Superkamiokande (SK) experiment 
indicates that $\nu_\mu$'s  oscillates to $\nu_\tau$'s
with a mass squared difference of $\approx 3\times 10^{-3}{\rm
eV^2}$~\cite{Fukuda:1998mi}, a result confirmed by K2K~\cite{Ahn:2002up}.  The simplest interpretation of this result is
that at least 
one of the neutrino mass eigenstates has a non-zero mass of $m\geq
5.5\times 10^{-2}{\rm eV}$, and as such provides the first 
glimpse of physics beyond the Standard Model (SM).  Recent
measurements of the solar neutrino spectrum at the Sudbury Neutrino Observatory (SNO) have 
confirmed the solar neutrino 
results of SK and found that $\nu_e$'s from the sun are
oscillating to the $\nu_\mu-\nu_\tau$
sub-system~\cite{Neubauer:2001pf, Ahmad:2002jz, Ahmad:2002ka}.  The
corresponding mass squared difference is smaller.  

The standard treatment of neutrino mixing has three active flavour
eigenstates mixed with three mass eigenstates.  This mixing is
described by the rotation of one basis into the other by the angles
$\theta_{13}$, $\theta_{12}$ and $\theta_{23}$ and a complex phase.
The atmospheric data constrains $\sin\theta_{23}$ to be almost maximal
with a mass squared difference of $\V\delta m_{13}\V\approx 3\times 10^{-3}{\rm
eV^2}$, while the solar neutrino data also constrains $\theta_{12}$ to be
large  and has $\V\delta m_{12}\V\leq 3\times 10^{-4}{\rm
eV^2}$~\cite{Gonzalez-Garcia:1999aj}.  These ``large mixing angle
results''
for the solar neutrino mixing were recently confirmed by the
terrestrial Kamland experiment~\cite{Eguchi:2002dm}.
  Furthermore $\theta_{13}$ is
bound by reactor 
data such that $\sin^2\theta_{13} \leq 0.1$~\cite{Apollonio:1999ae}.

The signs of the mass differences are still unknown, while information on
possible complex CP violating phases is extremely scant.  In order to overcome
this there have been a number new experimental facilities
proposed such as the 
the JHF neutrino beam,  and  others based on a muon storage
ring.  All provide a high intensity neutrino beam over a wide range of base
lengths and
energies~\cite{Itow:2001ee, Finley:2000cn, Ankenbrandt:1999as}.

Alongside the neutrino oscillation industry there has been an ongoing
and long established program of precision experiments aimed at measuring
the properties of the weak interaction.  Measurement of the Tritium
beta decay spectrum end-point is one such program while the
determination of the Michel parameters in the $\mu$ and $\tau$-decay
spectra is
another~\cite{Fritschi:1986kd,Robertson:1991vn, Belesev:1995sb,
Lobashev:1999tp, Weinheimer:1993pd, Rouge:2000um, Fetscher:2000th}. 
Originally the focus of these experiments was to establish
the Lorentz structure of the weak interaction, that is to decide
whether it is Vector minus
Axial-Vector (V-A) or some other combination of Scalar and Tensor
operators~\cite{Jackson:1957}.  Now that the V-A structure has been
established  the focus has shifted to measuring the electron neutrino
mass, at least in the case of
the Tritium 
experiments~\cite{Weinheimer:1999tn}.  Not until very lately have the possible
effects of neutrino 
mixing on these experiments been considered~\cite{McKellar:2001hj,
Stephenson:2000mw, Stephenson:1998cx, Farzan:2001cj}. 

In this paper the  converse scenario is examined, that is the impact of
non-standard interactions, such as additional Lorentz structures in the
weak interaction, on future neutrino oscillation experiments.  The aim of this work is to derive a formalism that will
describe the production, propagation and detection of a massive
neutrino with general Lorentz structures present at each of these
stages.  In effect we aim to incorporate the couplings that comprise
the Michel parameters, and the $\beta$-decay parameters of
Ref.~\cite{Jackson:1957} into the
description of neutrino oscillations.
 Similar calculations have focussed
on specific extensions to the Standard Model, or have been performed
exclusive of one or more of the production, propagation or detection
process. The correspondence between these previous calculations and
this work is expounded. 

\section{Background to NSI}
  In
the SM weak processes are mediated via the exchange
of a charged vector boson, and are  successfully described at low
energies by a $V-A$ 
current-current interaction with an effective coupling strength $G_F$.
Presently the SM has withstood all experimental tests  with the
exception of the recently observed apparent non-zero mass of the
neutrino.  Despite this there are many theories that seek to extend
the SM, often motivated by the desire to restore a
broken symmetry of the SM or to unify the strong force with the
electroweak force, or even to unite all four of the fundamental
forces under one theory.  Usually this unification occurs at a higher
mass scale than the SM, that is the bosons that mediate the new
interactions have a mass larger than the W and Z bosons of the
electroweak theory. The new interactions will manifest themselves as
current-current theories at low energies in the same way as the SM
charged current reduces to Fermi's effective field theory for energies
significantly less than the mass of the Z-boson.

If new physics is present at higher mass scales the low energy result will
be to induce new effective interactions with a coupling strength
weaker than the coupling of the dominant $V-A$ interaction unless the
fundamental coupling constant is anomolously large.  Absent a
specific model
there is no reason to expect that the new physics has the same
Lorentz structure as the weak interaction, hence new currents may be
a scalar $(S)$, pseudo-scalar $(P)$ or tensor $(T)$ in nature or comprise
of different combinations of vector and axial vector operators. Given
the dominance of the left-handed $V-A$ interaction we find it convenient to
cast these new interactions into forms with definite
handedness\footnote{To
 be completely
general one would need to also consider new V-A interactions in
addition to the standard interaction.  This has been the subject of
numerous investigations, see
Ref~\cite{Ota:2001pw}\cite{Johnson:1999ci}\cite{Huber:2002bi}.},
$L=V-A$, $R=V+A$, $S_L=S-P$ and  $S_R=S+P$. 

Assuming lepton universality the most general interaction Hamiltonians
for low energy, leptonic 
and semi-leptonic processes are given by
\bea
H_L &=& \sum_{\alpha\beta} g^{\alpha\beta}\bar{\psi}_l\Gamma_\beta
\psi_{\nu_l^{\alpha\beta}}\left(\bar{\psi}_{m}\Gamma_\alpha
\psi_{\nu_m^{\alpha\beta} }\right)^\dagger + h.c. ,\label{eq:leptham}
\eea
and,
\bea
H_{SL} &=& \sum_{\alpha\beta}
 G^{\alpha\beta}\bar{\psi}_l\Gamma_\beta \psi_{\nu_l^{\alpha\beta}}
\left(\bar{\psi}_q\Gamma_\alpha \psi_{q'}\right)^\dagger +
 h.c.\label{eq:semiham} 
\eea
respectively.
Where $\alpha/\beta= L,\ R,\ S_L,\ S_R$, while $\psi_{q}\  {\rm and}\ \
\psi_{q'}$ are quark fields, $\psi_l\ {\rm and}\ \psi_{m}$ are charged
lepton fields of flavour $l$ and $m$ respectively. The 
neutrino field is $\psi_{\nu_l^{\alpha\beta}}$, representing a
neutrino produced by an $(\alpha\beta)$ type interaction and associated
with a charged lepton of flavour $l$.  
The operators $\Gamma_\lambda$ are a combination of one of
the five bilinear covariants
\bea
\Gamma_L &=& \gamma_\nu(1-\gamma_5)\ ,\nn\\
\Gamma_R &=& \gamma_\nu(1+\gamma_5)\ ,\nn\\
\Gamma_{S_L} &=& (1-\gamma_5)\ ,\nn\\
\Gamma_{S_R} &=& (1+\gamma_5)\ .
\eea
In the SM only $\alpha=(L,R)$ and $\beta = L$ are present in the
semi-leptonic case with
$\psi_{\nu_l^{LL}}=\psi_{\nu_l^{RL}}$.  And for
the standard leptonic interaction only $\alpha=L$ and  $\beta = L$ are allowed.  

The fact that neutrinos oscillate indicates that the interaction
eigenstate is not the same as the mass eigenstate. Typically the two
are related via a unitary transformation
\bea
\psi_{\nu_f} = \sum_i U_{fi}\psi_{\nu_i}\ .
\eea 
Phenomenologically, there is no reason why a neutrino produced by one
 interaction needs to be coupled to the same combination
of mass eigenstates as a neutrino produced by another interaction.  So in
this work neutrino mixing is described by
\bea
\psi_{\nu^{\alpha\beta}_f} = \sum_i U_{fi}^{\alpha\beta}\psi_{\nu_i}\ .
\eea
Or equivalently each new interaction has its own unitary mixing matrix
which may be equivalent, or not, to the standard mixing matrix. 

The Eqs.~\ref{eq:leptham} \& \ref{eq:semiham}
describe the production and detection of a neutrino associated with a
charged lepton, in addition Eq.~\ref{eq:leptham} can describe the
interaction of a neutrino, via a charged current, with a background
medium of charged leptons.  This interaction leads to the well known
resonant enhancement of the neutrino mass, the MSW effect, crucial to
the solution of the solar neutrino
problem~\cite{Wolfenstein:1978ue, Mikheev:1986wj}.  In order to be used in
the derivation of the matter 
induced potential, to which we will return in a later section,
Eq.~\ref{eq:leptham} needs to be cast in the form of a neutral
current.  This is achieved by Fierz rearrangement (See for example
Ref.~\cite{Mohapatra:1998rq}) such that
\bea
-{\cal L}&=&\sum_{\alpha,\beta}g^{\alpha\beta}_{lf}
\left(\bar{\psi}_l\Gamma^\alpha\psi_{\nu_l}^{\alpha\beta}
\right)^\dagger\bar{\psi}_f\Gamma^\beta\psi_{\nu_f}^{\alpha\beta}\nn\\
&=&\sum_{\alpha,\beta}\sum_{ij}K^{\alpha\beta}_{(l)ij}\bar{\psi}_{\nu_j}\Gamma_\beta\psi_{\nu_i}\left(\bar{\psi}_{l}\Gamma^\alpha\psi_{l}\right)^\dagger
\label{eq:ncLagrangian}
\eea  
where
\bea
K_{(l)ij}^{LL} &=& -g^{LL}U^{LL}_{li}U^{LL}_{lj}, \nn\\
K_{(l)ij}^{RR} &=& -g^{RR}U^{RR}_{li}U^{RR}_{lj},   \nn\\
K_{(l)ij}^{RL} &=& {1\over 2}g^{S_LS_L}U^{S_LS_L}_{li}U^{S_LS_L}_{lj},  \nn\\
K_{(l)ij}^{LR}&=& {1\over 2}g^{S_RS_R}U^{S_RS_R}_{li}U^{S_RS_R}_{lj}, \nn\\
K_{(l)ij}^{S_LS_L} &=& {1\over 2}g^{S_RS_L} U^{S_RS_L}_{li}U^{S_RS_L}_{lj}
+ 3g^{T_RT_L}U^{T_RT_L}_{li}U^{T_RT_L}_{lj},\nn\\
K_{(l)ij}^{S_RS_R} &=& {1\over 2}g^{S_LS_R}U^{S_LS_R}_{li}U^{S_LS_R}_{lj} 
+ 3g^{T_LT_R}U^{T_LT_R}_{li}U^{T_LT_R}_{lj},\nn\\
K_{(l)ij}^{S_RS_L} &=& 2g^{RL}U^{RL}_{li}U^{RL}_{lj},\nn\\
K_{(l)ij}^{S_LS_R} &=& 2g^{LR}U^{LR}_{li}U^{LR}_{lj},\nn\\
K_{(l)ij}^{T_LT_L} &=& {1\over 4}g^{S_RS_L}U^{S_RS_L}_{li}U^{S_RS_L}_{lj} 
- {1\over 2}g^{T_RT_L}U^{T_RT_L}_{li}U^{T_RT_L}_{lj}, \nn\\
K_{(l)ij}^{T_RT_R} &=& {1\over 4}g^{S_LS_R} U^{S_LS_R}_{li}U^{S_LS_R}_{lj}
- {1\over
2}g^{T_LT_R}U^{T_LT_R}_{li}U^{T_LT_R}_{lj}\label{eq:propLagrangian} \
. 
\eea

This potential has been written in the mass basis so that
\Eq\ref{eq:ncLagrangian} actually describes the interaction of a
mass eigenstate, rather than a flavour
eigenstate, as is usually presented.  The effective 
coupling constant $K_{(l)ij}^{\alpha\beta}$ is dependent on the type
of leptons in the background as it contains the various mixing
elements specific to the lepton flavor. We do not included a neutral
current term for two reasons. Firstly, the standard
neutral current  contributes only to the absolute value of the
effective neutrino masses and not to the mass differences or
mixing angles.  And
secondly, any non-standard neutral current is highly
constrained by experimental data from atomic physics and LEP.
  
In this work we apply the
formalism being developed presently 
to terrestrial experiments, hence a background of charged
fermions only is considered. In the general case one would also need to
consider the effects of forward scattering off a neutrino background.
This has important consequences in early universe physics, described
 for example in
\cite{Cardall:1999bz, Thomson:1991hk, McKellar:1994ja}.

As alluded to earlier the Lorentz structure of the various currents in
the effective Lagrangians presented above may be generated by the
presence of physics beyond the SM. One prominent example is the low
energy effects of Supersymmetry~\cite{Mohapatra:1986uf}.  In particular
R-parity violating modes lead to charge-changing scalar
currents~\cite{Datta:2000ci}.  In the model discussed in
Ref.~\cite{Datta:2000ci}  stau mediated lepton number
violating decay, $\mu^-\to e^-\nu_e\bar{\nu}_\mu$, is examined. For
instance the
correspondence 
between the Lagrangian that produces this decay and
Eq.~\ref{eq:leptham} may be found by setting $\psi_{\nu^{S_LS_L}_e}\equiv
\psi_{\nu^{LL}_\mu}$ and
$\psi_{\nu^{S_LS_L}_\mu}\equiv\psi_{\nu^{LL}_e}$ and if for   the
supersymmetric coupling the following substitution is made 
$\lambda_{132}\lambda_{231}/(s-\tilde{m}_\tau^2)\equiv G^{S_LS_L}$,
where $\tilde{m}_\tau$ is the stau mass and $\lambda_{132}$ and
$\lambda_{231}$ are coupling constants.
Supersymmetry aside, right-handed vector currents can be motivated by
left-right 
symmetric models where a heavier W boson couples to right-handed
neutrinos, while scalar, vector and tensor currents may arise from
Leptoquark theories~\cite{Pati:1974yy}.

\section{Formalism---Scattering theory}
Spectral measurements such as the Tritium and
Michel parameter experiments yield little information about the mixing
elements $ U_{fi}^{\alpha\beta}$, hence it is natural to look to
oscillation experiments to access these parameters.  To do this a formalism 
that describes the oscillation of a neutrino with information about both
the production and detection processes is needed.
To do this we  treat the whole production, propagation and
detection process as a single scattering event.  In this scenario
the neutrino and charged vector bosons at the production and detection
sites are treated as unobserved intermediate states.  The oscillation
phenomena arises through the interference between these scattering diagrams.  
Now if NSI are present additional diagrams with the
non-standard boson at the production and/or detection site need to be
included.

Field theoretic (scattering theory) calculations of neutrino
oscillations have been performed by a number of
authors~\cite{Beuthe:2001rc, Grimus:1998uh, Grimus:1996av,
Kiers:1998pe, Giunti:1993se}.  For a  complete list of references
and a thorough review see the work by Beuthe,
Ref.~\cite{Beuthe:2001rc}. In this note the formalism developed by  
Cardall \& Chung, Ref.~\cite{Cardall:1999bz} is adapted to cater for NSI.  

To keep the notation transparent the formalism will be developed for a
specific example, that
of $\mu^+$-decay at the source, electron neutrino\footnote{The term electron
neutrino is used loosely here since neutrinos produced via different
interactions and associated with the 
decay of muons are  are not
necessarily the same. This is due to the various mixing matrices
associated with the different interactions. When it matters more
specific language will be used.} oscillation and subsequent detection of a negatively charge muon at
the detector. The scattering amplitude is given by the time ordered
product of interaction Hamiltonians
\bea
A & = &  <\bar{\nu}_\mu e^+ X \mu^-\V T[\int d^4x \int d^4 y H_L(x)H_{SL}(y)]\V\mu^+
N>\nn\\
 & = & \int^\infty_{-\infty} dx^0\int^\infty_{-\infty} dy^0\int_{V_S}\!\!
d^3x\int_{V_D}\!\! d^3y{1\over\sqrt{(2\pi)^6V_S^3V_D^3}}
\nn\\ &\times &
{\exp[-i(p_{\mu^+}-p_e-p_{\bar{\nu}})\cdot
x]\over\sqrt{(2E_{\mu^+})(2E_e)(2E_{\bar{\nu}})}}  
M_{fi}(x,y\V p_i,p_f,q){\exp[-i(p_N-p_{\mu^{-}}-p_X)\cdot y]
\over\sqrt{(2E_N)(2E_{\mu^{-}})(2E_X)}}\ ,
\eea
where $E_{\mu^+}$, $E_e$ and $E_{\bar{\nu}}$, are the energies of the
anti-muon, 
electron and the muon anti-neutrino (not participating in the
oscillation measurement) with corresponding four momenta
$p_{\mu^+}$, $p_e$ and $p_{\bar{\nu}}$.  The reaction at the source is
evaluated at a 
space-time point $x$ in a volume $V_S$. The energies of the
particles at the detector are $E_N$, $E_{\mu^{-}}$, $E_X$ with associated
four-momenta $p_N$, $p_{\mu^{-}}$, $p_X$.  The reaction at the detector is
evaluated at a space-time point $y$ in a volume $V_D$.  Furthermore the
matrix element 
$M_{fi}(x,y\V p_i,p_f,q)$ is given by
\bea
M_{fi}(x,y\V p_i,p_f,q) & = &
\sum_{ij}\sum_{\alpha\beta}\sum_{\lambda\sigma}g^{\alpha\beta}G^{\lambda\sigma}
U^{\alpha\beta}_{\mu i}U^{\lambda\sigma}_{\mu j}J^{\lambda}(p_N,p_X)^\dagger\nn\\
&\times& \left(\bar{u}(p_{\mu^{-}})\Gamma_\sigma G^{ji}(x,y)
\gamma^0\Gamma_\alpha^\dagger\gamma^0 v(p_e)\right)\cdot\left(\bar{v}(p_{\mu^+})\Gamma_\beta
v(p_{\bar{\nu}})\right)\ .\label{eq:matrixele}
\eea
Here $p_i$ and $p_f$ are the momenta of the particles in the initial
and final state respectively, $q$ is the momentum of the oscillating
neutrino while $J^{\lambda}(p_N,p_X)$ is the
nuclear matrix element for the transition operator $\Gamma_\lambda$.
The sums are over neutrino vacuum masses, source and detector interactions
respectively.  The function $ G^{ij}(x,y)$ is the neutrino propagator or
Green's function.  In vacuum, $ G^{ij}(x,y)\to G^{ii}(x,y)$ and is
given by the standard Dirac propagator.

The starting point in the evaluation of $G^{ij}(x,y)$ is the equation of motion for a propagating neutrino
from which the equation satisfied by the neutrino
Green's function can be found, 
\bea
(i\delslash-M-V) G(y,x)=\delta^4(y-x)\ ,\label{eq:eom}
\eea
where $V$ is the propagation potential derived from
Eq.~\ref{eq:ncLagrangian} and M is the mass matrix,
note the mass indices have been suppressed as we will be concerned
with $G(y,x)$ in spinor space for the moment.
 Due to the chiral nature of the operators
$\Gamma_\sigma$ and  $\Gamma_\alpha^\dagger$ it is convenient to
write the Green's function in chiral blocks
\bea
G(x,y)= \left(\begin{array}{cc} G_{LL}(x,y) & G_{LR}(x,y) \\ G_{RL}(x,y) &
G_{RR}(x,y)\end{array}\right)\label{eq:chiralGreen}\ ,
\eea
where $G_{XY}$ is the element projected out by $P_X G(x,y) P_Y$ where
$P_X = P_{R/L}= \frac{1}{2}(1\pm\gamma_5)$.   Similarly the operator in Eq.~\ref{eq:eom} in
chiral form can be written as
\bea
(i\delslash-M-V) = \left(\begin{array}{cc} - M_1   &
\sigma\cdot(i\partial - V^{RR}) \\
\bar{\sigma}\cdot(i\partial-
V^{LL})  & -M_2  \end{array}\right)\label{eq:chiralOpp}\ ,
\eea
where
\bea
M_1 &=& M+S^{S_RS_L}
\eea
and
\bea
M_2 &=& M+S^{S_LS_R}\ , 
\eea
with
\bea
S^{S_RS_L} = \sum_l\int {d^3p\over (2\pi)^3}\rho_l(\vec{p}_l) {m_l\over
E_l} (K^{S_RS_L}+K^{S_LS_L})\ ,\label{eq:scalarpot}\\
S^{S_LS_R} = \sum_l\int {d^3p\over (2\pi)^3}\rho_l(\vec{p}_l) {m_l\over
E_l} (K^{S_LS_R}+K^{S_RS_R})\ ,\\
(V^{LL})^\mu = \sum_l\int {d^3p\over (2\pi)^3}\rho_l(\vec{p}_l) {p_l^\mu\over
E_l} (K^{LL}+K^{RL})\ ,\label{eq:leftpot}\\
(V^{RR})^\mu = \sum_l\int {d^3p\over (2\pi)^3}\rho_l(\vec{p}_l) {p_l^\mu\over
E_l} (K^{RR}+K^{LR})\ .\label{eq:vecpot}
\eea
The sums in Eqs.~\ref{eq:scalarpot}-\ref{eq:vecpot} are over lepton
species with mass $m_l$ distributed according to $\rho_l(\vec{p}_l)$
in the background 
medium.  In general the Green's function may 
be expressed in any basis, however we find it convenient to do so in the basis
defined by the vacuum mass eigenstates.  In this basis the matrix M is
diagonal, while the matrices $S^{S_RS_L}$, $S^{S_LS_R}$,
$(V^{LL})^\mu$ and $(V^{RR})^\mu$ are in general not diagonal.  All other
operators are proportional to the unit in this basis assuming the
potential $V$ does not vary in space or time. 

Using \Eq\ref{eq:chiralGreen} \& \Eq\ref{eq:chiralOpp}
two sets of two equations can be found and used to solve for each component
of the Green's function.  For $G_{LR}$ and $G_{RR}$;
\bea
\delta^4(x_2-x_1)&=&-M_2G_{RR}+\left[i(\partial_0
-\vec{\sigma}\cdot\vec{\nabla})-  
V^{LL}\cdot\bar{\sigma}\right]G_{LR}\label{eq:1firstset}\ ,\\
0&=&-M_1G_{LR}+\left[i(\partial_0
+\vec{\sigma}\cdot\vec{\nabla})-  
V^{RR}\cdot{\sigma}\right]G_{RR}\label{eq:2firstset}\ ,
\eea
and for  $G_{LL}$ and $G_{RL}$;
\bea
\delta^4(x_2-x_1)&=&-M_1G_{LL}+\left[i(\partial_0
+\vec{\sigma}\cdot\vec{\nabla})-  
V^{RR}\cdot{\sigma}\right]G_{RL}\ ,\\
0&=&-M_2G_{RL}+\left[i(\partial_0
-\vec{\sigma}\cdot\vec{\nabla})-  
V^{LL}\cdot\bar{\sigma}\right]G_{LL}\ .\label{eq:2secondset}
\eea 
For the analysis that is to follow in the next section only the first
set of equations will need to be solved.  This is done following Cardall
\& Chung, Ref.~\cite{Cardall:1999bz}, by defining
\bea
J(x_2,x_1)= M_1^{-1}G_{RR}(x_2,x_1),\label{eq:J}
\eea 
which reduces the problem to solving
\bea
\left(-M_2M_1+ \left[i\bar{\sigma}.\partial
-v_L\cdot\bar{\sigma} \right]\left[i\sigma\cdot\partial-
v_R\cdot\sigma\right]\right)J(x,y)=\delta^4(y-x)\ ,
\eea
where $v_R= M_1^{-1} V^{RR} M_1$ and for brevity $v_L=V^{LL}$.  Now the
delta function and $J(y,x)$ are expanded into momentum space via a
Fourier transform where
we find that
\bea
\bigl(-M_2M_1+ \left[(q_0+\vec{\sigma}\cdot\vec{q})
-v_L\cdot\bar{\sigma} \right]\left[(q_0-\vec{\sigma}\cdot\vec{q})  -
v_R\cdot\sigma\right]\bigr)J(q_0,\vec{q})=1\ .
\eea
Where $q_0$ and $\vec{q}$ are the unobserved neutrino energy and
momenta.
The task is now to find the elements of $J(q_0,\vec{q})$ in terms of
its reciprocal matrix $R(q_0,\vec{q}) =J(q_0,\vec{q})^{-1}$.
 Recalling that $J(q_0,\vec{q})$ is a
matrix in the $2\times 2$ space of the Pauli spinors, one can make a
substantial 
simplification by choosing the direction of neutrino propagation to
coincide with third spatial coordinate.  For a non-polarised,
non-relativistic  background
of leptons we find that:
\bea
R(q_0,\vec{q})_{11}&=& q_0^2-\V\vec{q}\V^2-M_2M_1 -\left[q_0
v_R^0 + v_R^0\V\vec{q}\V\right]\nn\\
&-& \left[q_0 v_L^0-v_L^0\V\vec{q}\V\right]+ v_L^0v_R^0\ , \\
R(q_0,\V \vec{q}\V)_{12}&=&0\ ,\\
R(q_0,\V \vec{q}\V)_{21}&=&0\ ,\\
R(q_0,\V \vec{q}\V)_{22}&=& q_0^2-\V\vec{q}\V^2-M_2M_1 -\left[q_0
v_R^0 - v_R^0\V\vec{q}\V\right]\nn\\
&-& \left[q_0 v_L^0+v_L^0\V\vec{q}\V\right]+ v_L^0v_R^0\ ,
\eea 
where the subscripts correspond to the elements of the matrix $R(q_0,\vec{q})$.

To perform the Fourier transform back to coordinate space the pole
structure of $J(q_0,\vec{q})$ must be established.  This is
accomplished by defining the unitary matrices $\tilde{U}_R$ and
$\tilde{U}_L$ which diagonalize, in flavour space, $R_{11}$ and
$R_{22}$ respectively.  In the ultra-relativistic limit where
$\V\vec{q}\V\to q_0$,  the components of   $J(x_2,x_1)$ are 
\bea
J(\vec{y},\vec{x})_{11} & =& \int {dq_0\over 2(2\pi)^2}
{e^{-iq_0(y^0-x^0)}\over 
\V\vec{y}-\vec{x}\V}\sum_K e^{-iq_0\V\vec{y}-\vec{x}\V}
\tilde{U}_{iK}^{R} \tilde{U}^R_{jK} e^{i{m_{RK}^2\over
2q_0}\V\vec{y}-\vec{x}\V}\label{eq:jright} \\ 
J(\vec{y},\vec{x})_{22} & =& \int {dq_0\over 2(2\pi)^2}
{e^{-iq_0(y^0-x^0)}\over 
\V \vec{y} - \vec{x}\V}\sum_K e^{-iq_0\V \vec{y}-\vec{x}\V }
\tilde{U}_{iK}^{L} \tilde{U}^L_{jK} e^{i{m_{LK}^2\over
2q_0}\V\vec{y}-\vec{x}\V}\label{eq:jleft} \ . 
\eea
The effective masses $m_{RK}^2$ and $m_{LK}^2$ are the $K$ eigenvalues of the
 matrices
 $-M_2M_1+v^0m^0-2q_0 m^0$ and $-M_2M_1+ v^0m^0-2q_0 v^0$ respectively.

The components of the Green's function, $G_{RR}$ and $G_{LR}$, are found by
substitution into Eq.~\ref{eq:J} and then Eq.~\ref{eq:1firstset}.  In
the matrix element, Eq.~\ref{eq:matrixele}, the sums over vacuum
masses are carried out such that 
\bea
\sum_{i}\sum_{K}U^{\alpha\beta}_{\mu i}U^R_{iK}\to
\sum_KU^{\alpha\beta R}_{\mu K}
\eea
where the sum is now over effective masses.

  The
scattering amplitude is found by localising the
particles at the source and detector to boxes of volume $V_S$ and
$V_D$ and by expanding $\vec{x}$ and $\vec{y}$ about the
source and detection
points $\vec{x}_S$ and $\vec{y}_D$ respectively~\cite{Cardall:1999bz}. 
Finally we find for the square of the scattering amplitude,
\bea
{\cal A}_L{\cal A}_K^* &=& T {1\over {V_S}^{2}}{1\over {V_D}^{2}}  \left[(2E_{\mu^+})(2E_e)(2E_{\bar{\nu}})\right]^{-{1}}
\left[(2E_N)(2E_X)(2E_{\mu^-})\right]^{-{1}}\nn\\
&\times & (2\pi)^4\delta^4(p_{\mu^+} -p_e-p_{\bar{\nu}}-q)(2\pi)^3\delta^3(\vec{p}_N + \vec{q} -\vec{p}_{\mu^-} -\vec{p}_X) \nn\\ 
&\times &  
\left({1\over 4\pi\V \vec{L}\V}\right)^2\exp [i\delta
m_{ji}^2\V\vec{L}\V] 
({\cal M}_{\mu^+}^L{\cal M}_{\mu^+}^{K*})({\cal M}_{\mu^{-}}^{L}{\cal M}_{\mu^{-}}^{K*}) \ .
\eea 
To find a per source particle per detector particle event rate
integrate over the final state phase space (factor of $Vd^3p/(2\pi)^3$
for each particle in the final state) and divide by $T$ after
interpreting one of the energy delta functions as characterising the
time of the scattering process;
\bea
d\Gamma(E_q) & = & \int \sum_{KL} {\exp(i\delta m_{KL}^2/2E_q \V\vec{L}\V)\over
\V\vec{L}\V^2} {d^2N_\nu\over dE_q 
d\Omega_q}\Biggr|_{KL}\sigma(\nu N)_{KL}dE_q\ ,
\eea
The quantity ${d^2N_\nu/ dE_q d\Omega_q}$ is
interpreted as
the neutrino flux;
\bea
{d^2N_\nu\over dE_q d\Omega_q}\Biggr|_{KL} & =& \int {d^3p_e\over
(2\pi)^3}{d^3p_{\bar{\nu}}\over (2\pi)^3}{(2\pi)^4\delta^4(p_{\mu^+}
-p_e-p_{\bar{\nu}}-q) E_q^2 {\cal M}_{\mu^+}^{L*}{\cal M}_{\mu^+}^{K}\over
(2\pi)^3(2E_{\mu^+})(2E_q)(2E_{\bar{\nu}})(2E_e)}\ ,  
\eea
where the $(KL)$ dependence arises from the various mixing elements in
matter.
The $(\nu N)$ cross section also has a $(KL)$ dependence and is
given by
\bea
\sigma(\nu N)_{KL} & = & \int {d^3p_X\over
(2\pi)^3}{d^3p_{\mu^-}\over (2\pi)^3}{(2\pi)^4\delta^4(p_N +q
-p_X-p_{\mu^-}){\cal M}_{\mu^{-}}^L{\cal M}_{\mu^{-}}^{K*}  \over
(2E_q)(2E_N)(2E_X)(2E_{\mu^{-}})}\ .
\eea
The mixing elements are all contained in the squared matrix elements 
and are not in general able to be extracted and interpreted as
an oscillation probability, except in the case where only one type of
interaction is responsible for source and detection processes, or if
only one type of mixing matrix exists, i.e.~$U^{\alpha\beta}\equiv
U^{\gamma\delta}$. 

The event rate is obtained by integrating over the source and detector
particle distributions:
\bea
\frac{dN}{dE_q} = \int d^3x_S \int {d^3p_{\mu^+}\over (2\pi)^3}
f_\mu(\vec{p}_{\mu^+},\vec{x}_S) \int d^3y_D\int {d^3p_N\over (2\pi)^3}
f_N(\vec{p}_N,\vec{y}_D) \frac{d\Gamma}{dE_q}\ .\label{eq:totevent}
\eea

This event rate, with Eqs.~\ref{eq:2firstset} and \ref{eq:2secondset},
is the first main result of this paper.  It is an expression
incorporating neutrino oscillations in matter, and after following the
same procedure just outlined to solve for $G_{LR}$ \& $G_{LL}$,
allows for arbitrary Lorentz couplings.  There are three main points
of difference between this result and the standard result.  Firstly,
in vacuum there is the possibility of observing flavour violation as a
result of the intermediate neutrino even if the neutrino mass differences are
zero. This requires that the non-standard mixing matrix not be related
trivially to the standard matrix. In this scenario  no oscillatory phase will
develop.  Secondly, in matter, the addition of NSI allows for an
oscillatory phase even if the neutrino masses vanish.  Again the
non-standard mixing matrix must not be trivially related to the
standard matrix.  These results were first noted by Bergman~{\it et.~al.}~in 
Ref.~\cite{Bergmann:1999rz} and subsequently investigated by
Huber~{\it et.~al.}~in 
Ref.~\cite{Huber:2001zw}.  They found that NSI
could only act as a perturbation to the solution of the solar and
atmospheric problems rather than a complete explanation.  And thirdly,
we note that the presence of a right-handed potential $V^{RR}$
connects a left-handed neutrino state to a right handed state, even in
the ultra-relativistic limit.  This is akin to the spin flip caused by
a non-zero neutrino magnetic moment in a strong field.  We do not
pursue this here, however we note that the presence of this potential
in a dense medium will require the equations of motion for
right-handed and left-handed neutrinos  be solved simultaneously,
perhaps using a density matrix approach.   Finally we highlight the
fact that if the right-handed potential $v_R$ is zero then the only
non-zero component of the Greens function is $G_{22}$ in the
ultra-relativistic limit.  Thus the
connection to right-handed states vanishes and the only required
diagonalization matrix is $\tilde{U}^{L}$.

\section{Considerations for a neutrino beam}
The formalism developed in the previous section can now be applied to
a generic neutrino factory scenario.  In this section no
attempt is made at a full simulation of this type of experiment,
A full simulation would involve a discussion of detector properties,
efficiencies, energy resolutions, cuts, backgrounds and the like;
rather we attempt to define regions of interesting parameter space
and achieve a 
quantitative understanding of the effects of NSI. As we
have seen, in the relativistic limit, the event rate for a particular
experiment can be obtained by evaluating the neutrino flux,
${d^2N_\nu/ dE_q  
d\Omega_q}|_{KL}$, containing the effects of NSI at the source, the
production cross
section, $\sigma(\nu n)_{KL}$, accounting for NSI at the detector, and
the phase factor ${\exp(i\delta m_{KL}^2\V\vec{L}\V/2E_q )/ 
\V\vec{L}\V^2}$ dealing with the oscillatory phase.  In this paper we
present the results of two investigations.  Firstly the vacuum
propagation of a neutrino produced at the source via $\mu^-$-decay and
subsequent production of a $\tau^-$ at the detector, that is
$\nu_\mu\to \nu_\tau$ oscillation.  Secondly we examine the
case of a neutrino produced via 
$\mu^+$-decay, interacting with the electrons in the background
medium, and producing a $\mu^-$ in the detector, or $\nu_e\to\nu_\mu$
oscillation. In both cases this is
performed for the simple case of one non-standard scalar coupling,
$(\alpha,\beta)=(S_L,S_L)$, in addition to the standard
interaction in
Eqs.~\ref{eq:leptham} \& \ref{eq:semiham}.  This particular
non-standard coupling is chosen since it is the only interaction other
than the standard one that does not require the existence of
right-handed neutrinos.  The strength of the coupling for both the
leptonic and semi-leptonic interaction is taken to be
$g^{S_LS_L}=G^{S_LS_L}=0.01\ G^{LL}$, within the current experimental
upper bounds~\cite{Herczeg:2001vk, Herczeg:1996qn}.

The expression for the neutrino flux resulting from the decay of a
$\mu^-$ can be calculated using standard trace techniques, in general
one will find two new terms as a result of the scalar interaction, an
interference term between the standard and non-standard interaction
of order $\V g^{S_LS_L}\V$ and a pure scalar term of order $\V
g^{S_LS_L}\V^2$. The interference term 
reflects the right-handed admixture, proportional to
$\frac{m_\mu}{E_\mu}$, in the left-handed wavefunction of the
$\mu^-$~\cite{Kayser:1989iu}.  The chiral structure of the scalar
interaction being considered requires a right-handed $\mu^-$, while
the standard weak interaction always proceeds via a left-handed
$\mu^-$.  Since the process by which muons will be produced at a
neutrino factory will be dominated by the standard weak interaction,
e.g. $\pi$-decay, any contribution from  NSI will be doubley
suppressed, once by the production process and once by the decay
process.  For this reason we neglect the contribution from NSI at the
source.  However the expression for the neutrino flux with the
additional NSI terms is recorded here for completeness.  In the rest frame
of the muon, neglecting all lepton masses we find that
\bea
{d^2N_\nu\over dx d\Omega_q}\Biggr|_{ij} & = & \omega_\mu {2x^2\over
4\pi}\biggl[U^{LL}_{\mu j} U^{LL}_{\mu i}(3-2x) + 3\rho^2
U^{S_LS_L}_{\mu j}U^{S_LS_L}_{\mu i}(1-x)\biggr] \ ,
\eea
where $x=2m_\mu/E_\nu$ and $\rho= G^{LL}/G^{S_LS_L}$.

For energies
above the tau 
threshold of $\sim 5 {\rm GeV}$ the dominant reaction mechanism at the
detector is 
deep inelastic 
scattering producing many particles in the final state in addition to
the tau, denoted
collectively as $X$.
The $(\nu_\tau N)$ cross section can be derived using a parton model
of the nucleon.  In the standard treatment the lepton and parton
masses are typically assumed to be zero,  this is a good approximation in most
circumstances however its
validity for this application is 
questionable.  This is due in part to the considerable mass of the $\tau$
and also to the fact that certain mixing patterns may act to pick out
the interference term in the cross section which would otherwise vanish in
the massless limit as we shall show.  A treatment using massive partons
has been given by Aivazis~{\it et.~al.}~in Ref~\cite{Aivazis:1994kh},
and the
non-standard cross section is derived using this formalism.

The general expression for the cross section with a vanishing neutrino
mass is
\bea
d\sigma &=&{2\pi Q^2 \over \Delta(s,0,M^2)}\biggl(\V G^{LL}\V^2
L^{\alpha\beta}W_{\alpha\beta}\nn\\ 
&+& 2G^{LL}G^{S_LS_L}L^\alpha W_\alpha
+ \V G^{S_LS_L}\V^2 LW\biggr)d\Gamma\ ,\label{eq:xsec}
\eea
where $\Delta(s,0,M^2)$ is a kinematic function of the Mandelstam
variable $s$ and the nucleon mass $M$, while we have retained a
non-zero $\tau$-mass we have neglected the small neutrino mass.  The
final state phase space is  
represented by $d\Gamma$.  The complete
evaluation of \Eq(\ref{eq:xsec}) is lengthy but involves only standard
trace algebra, here we just state the results.  
The cross section in terms of the Bjorken scaling variables
$x$ and $y$ is written as
\bea
{d\sigma\over dx dy}\Biggr|_{ij} &=& U^{LL}_{\tau i}U^{LL}_{\tau
j}\left[{d\sigma\over dx dy}\right]_{LL} + U^{S_LS_L}_{\tau
j}U^{S_LS_L}_{\tau i} \left[{d\sigma\over dx dy}\right]_{SS} \nn\\
&+& {1\over 2}\left( U^{LL}_{\tau
j}U^{S_LS_L}_{\tau i}+ U^{S_LS_L}_{\tau j}U^{LL}_{\tau
i}\right)\left[{d\sigma\over dx dy}\right]_{SL}
\eea
where the first two terms arise from vector and scalar interactions
respectively and the final term arises from the interference between
the two. 
In the limit where all terms of order $m_i^2/Q^2$  or
greater can be ignored (where
in this context $m_i$ corresponds to the parton masses) we find
\bea
\left[{d\sigma\over dx dy}\right]_{SS} \approx \V G^{S_LS_L}\V^2 {ME_\nu
\over 2\pi}  \left(xy^2 + \frac{m_\tau^2 y}{2ME_\nu}\right)F_1(x)\label{eq:SS},
\eea
\bea
\left[{d\sigma\over dx dy}\right]_{SL} \approx
G^{S_LS_L}G^{LL}\frac{ME_\nu}{2\pi}m_\tau\left(\frac{1}{E_\nu} -
(\frac{xy}{E_\nu} +\frac{m_\tau^2}{2M E_\nu^2})\right)F_1(x),\label{eq:SL}
\eea
The  term, $\left[{d\sigma/ dx dy}\right]_{LL}$, not shown here is the standard results and was derived in
Ref~\cite{Albright:1975ts}. 
The structure functions, $F_1$ through $F_5$ are measured quantities fitted
to a functional form, we use the same form as is used in
Ref~\cite{Hall:1998ey}.  In the limit that $m_1^2/Q^2\to 0$ the struck
parton mass becomes $m_1\to xM$ as noted in Ref~\cite{Leader:1996hm}.

With the assumption of an idealised source of neutrinos arising from
muon decay the charged current event rate can be written as
\bea
N_\tau &=& 6.023\times 10^{32}\frac{N_\mu M_{kt}}{E_\mu \V
\vec{L}\V^2}\nn\\
&\times&\int\left[\sum_{ij}\frac{d^2\rho_\nu^{*}}{dxd\Omega_q}\Biggr|_{ij}
 {\sigma_{\nu N}}|_{ij}
\exp(-i\frac{\delta m^2_{ij}\V\vec{L}\V}{2E_q})
\right] dE_q\ ,\label{eq:idealeventrate} 
\eea
where
$\omega_\mu{d^2\rho_\nu^{*}}/{dxd\Omega_q}={d^2N_\nu^{*}}/{dxd\Omega_q}$
is the neutrino distribution in the reference frame of 
the lab.  For high energy muons $x\to E_q/E_\mu$ and:
\bea
{d^2\rho_\nu^*\over dxd\Omega_q} & = & {1\over
\gamma^2(1-\beta\cos\alpha)^2}{d^2\rho_\nu\over dxd\Omega_q}\nn\\
&\approx & 4\gamma^2 {d^2\rho_\nu\over dxd\Omega}
\eea  
where $\alpha=0$ coincides with the beam direction, $\gamma =
E_\mu/m_\mu$ and $\beta = p_\mu/E_\mu$.  For 
high beam energies $\cos\alpha \sim 1$ and $\beta \sim 1-1/(2\gamma)$.  
The number of useful muon decays per year, $N_\mu$, is defined as the
integration over the source distribution times the $\mu$-decay rate. 
 The parameter $M_{kt}$ is the mass of the
detector in kilotons.  The numerical factor in
\Eq(\ref{eq:idealeventrate}) is the number of nucleons in the detector
per kiloton, we have assumed an ideal detector comprising $ 6.023\times
10^{32}M_{kt}$ non-relativistic nucleons.

\section{Vacuum oscillations---Two neutrinos.}\label{sec:vacosc}
We now study the effects of varying the non-standard mixing angle for
the $\nu_\mu\to \nu_\tau$ vacuum oscillation channel.  

The study is based on the philosophy that the atmospheric, solar
and Kamland neutrino data sets have accurately defined the leading oscillation
parameters.  That is  the LMA solution is accurate and 
the NSI are treated as perturbations to this solution.  This
philosophy is backed 
by the study of atmospheric neutrinos and NSI in Ref~\cite{Fornengo:2001pm}.
We examine a
simplified two neutrino system through the $\nu_\mu\to \nu_\tau$
oscillation channel.  As such  the
standard and non-standard mixing matrices are defined as
\bea
U^{LL} = \left(\begin{array}{cc} \cos\theta & \sin\theta \\
-\sin\theta & \cos\theta \end{array}\right)\ {\rm and}\  \ 
U^{S_LS_L} = \left(\begin{array}{cc} \cos\phi & \sin\phi \\
-\sin\phi & \cos\phi \end{array}\right)\ .
\eea
The standard mixing angle and mass difference given by the LMA
solution are $\sin^2(2\theta)\approx 1$ and $\delta m^2 \approx
2.5\times 10^{-3}\ {\rm eV^2}$.

The baselines over which it is proposed that oscillation experiments
will be conducted   
vary from a few hundred kilometres to several thousand.  For example
at the proposed Fermilab neutrino factory there are plans
to place detectors $732\ {\rm km}$ away at the Soudan detector and even
at the South Pole a distance of $11700\ {\rm km}$.  The energies of
the muon beam used to produce the neutrino flux are to be
optimised for the detection of CP violation in the neutrino sector.
Typically energies of $20-50\ {\rm GeV}$ are being studied. The great
advantage using a muon storage ring as a neutrino source is the
extremely high intensity of the resulting beam with $\sim 10^{21}$
neutrinos expected to be produced per year.

Other studies
have investigated the effects on oscillations of varying the base
length $\V\vec{L}\V$ 
and the beam energy $E_q$ with NSI present~\cite{Ota:2001pw}.  In any
case variations in the beam length will have little consequence for
the parameter range we will consider; that is for energies, mass
difference and base lengths where  $\delta m^2_{ij}\V\vec{L}\V/ 2E_q<<
1$.  
For the $\nu_\mu\to \nu_{\tau}$ oscillation channel in the absence of
NSI
\bea 
P_{\mu\tau} &\approx &
2(\cos\theta\sin\theta)^2-2(\cos\theta\sin\theta)^2\left(1-\frac{1}{2}
\left(\frac{\delta m^2_{ij}\V\vec{L}\V}{2E_q}\right)^2\right)\nn\\ 
&=&(\cos\theta\sin\theta)^2\left(\frac{\delta
m^2_{ij}\V\vec{L}\V}{2E_q}\right)^2.
\eea
When this is substituted into an event rate such as
\Eq(\ref{eq:idealeventrate}) the dependence on $\V\vec{L}\V$ is
removed. 

The sensitivity of a future neutrino factory to NSI is
investigated by defining a $\chi^2$ function which determines the required
detector mass and number of useful muon decays in order to claim new
physics.
The $\chi^2$ function assuming Gaussian statistics for a detector mass
and muon number of $N_\mu\cdot 
M_{det}=10^{21} {\rm kt/yr}$ is defined as
\bea
\chi^2_{NM21} = \sum_k {\V N_k^{SM} -N_k^{NSI}\V^2\over N_k^{SM}}\ , 
\eea 
where $N_k^{SM}$ is the expected number of $\tau$ producing charged
current events in energy bin $k$ in the absence of NSI and for this
detector mass and muon number
\bea
N_k^{SM} &=& 6.023\times 10^{32}\frac{N_\mu M_{kt}}{E_\mu \V
\vec{L}\V^2}\nn\\
&\times&\int^{E_k}_{E_{k-1}}\left[\sum_{ij}\frac{d^2\rho_\nu^{*}}{dxd\Omega_q}\Biggr|_{ij} 
 {d\sigma_{\nu N}}|_{ij}
\exp(-i\frac{\delta m^2_{ij}\V\vec{L}\V}{2E_q})
\right] dE_q\ .\label{eq:idealeventbin}
\eea
Here the energy bins are defined such that $E_{k-1}< E_q< E_k\ (k
=1,2,3,..,n)$.
Furthermore $N_k^{NSI}$ is the number of charged current events in the
$k^{th}$ energy bin expected with the NSI
present. We define a required 
detector mass-muon number unit, $NM_{rec}=1\times 10^{21}\ {\rm
kt/yr}$, to quantify the sensitivity to NSI for a given set of
parameters.  The constraint on $NM_{rec}$ is  
\bea
NM_{rec} > {\chi^2_{90\%}\over \chi^2_{NM21}},
\eea
where $\chi^2_{90\%}$ is the $\chi^2$ value at $90\% $ confidence
level and one detector mass-muon number unit defined as a function of
the number of degrees of freedom\footnote{The subtleties of performing
this kind of analysis have been examined in Ref.~\cite{Ota:2001pw}.}.  The sensitivity of $NM_{rec}$ to
the number of energy bins used, where the number of bins corresponds to the
number of degrees of freedom, was investigated by varying the number
of bins for various values 
of beam energy.  We found the result to be statistically stable for
ten bins or greater.

\begin{center}
\begin{figure}[p]
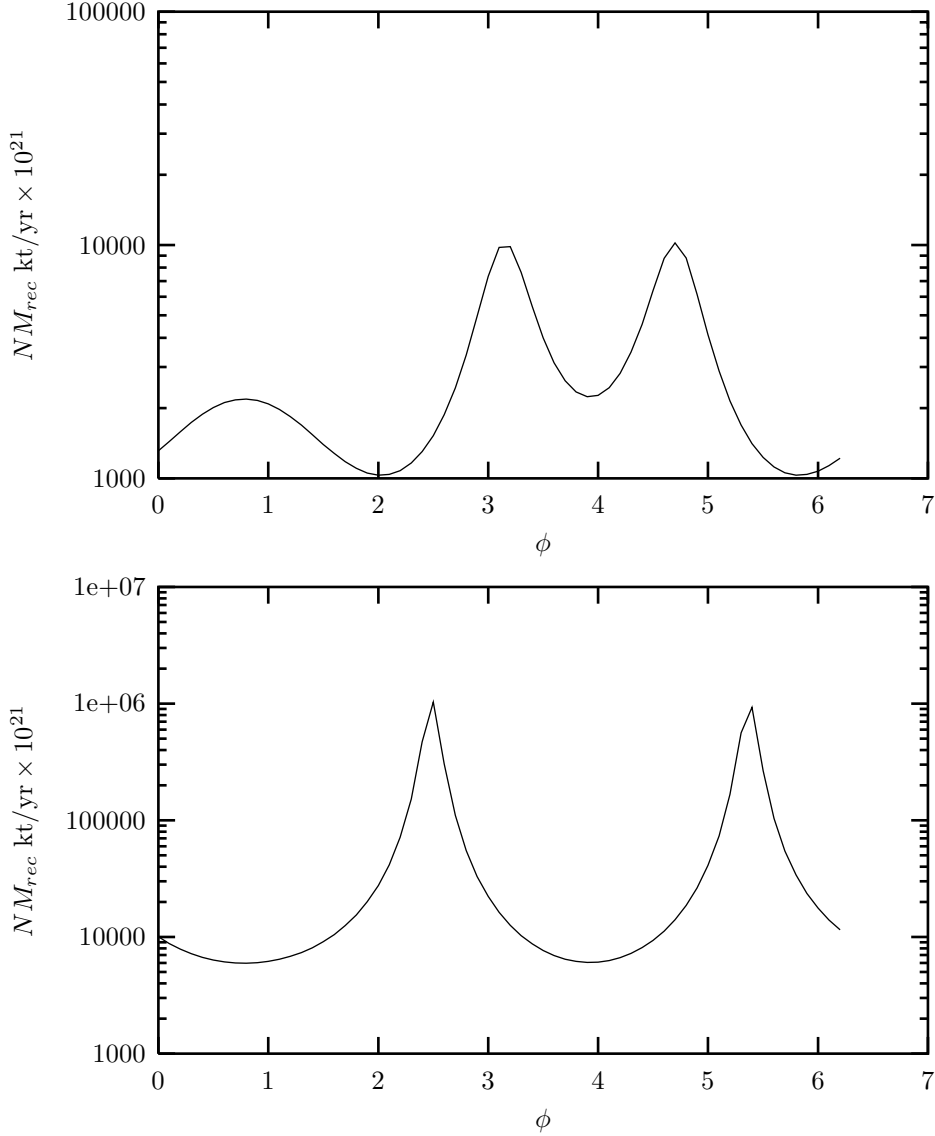

\centering
\input{phi.E50.ps}
\input{phi.E20.ps}
\caption{Top panel: The required number of detector mass-muon number
units $NM_{rec}$ with 
$G^{S_LS_L}=0.01G^{LL}$, $\delta 
m^2 = 0.0025 eV^2$, $L = 732km$.
Bottom panel: The required number of detector mass-muon number
units $NM_{rec}$ with $G^{S_LS_L}=0.01G^{LL}$, $\delta
m^2 = 0.0025 eV^2$, $L = 732km$. }
\label{fig:phi.E50.ps}
\end{figure}
\end{center}

 In
Fig.~\ref{fig:phi.E50.ps} the variation of $NM_{rec}$ with
respect to the non-standard mixing angle 
$\phi$ is shown.  
The first plot is for a beam energy of $E_\mu = 50\
{\rm GeV}$ , while the second is for a  beam energy of $E_\mu = 20\
{\rm GeV}$. Points of particular interest are $\phi=
\pi/4 ,\  5\pi/4$ and $\phi = 3\pi/4,\ 7\pi/4$.  The first case
corresponds to the situation $\nu_\mu^{S_LS_L}\equiv \nu_\mu^{LL}$ or
$U_{\mu i}^{S_LS_L}\equiv U_{\mu i}^{LL}$ up to a phase. For this case
$N M_{rec}$ is close to a maximum.  In the second case
$\nu_\mu^{S_LS_L}\equiv \nu_\tau^{LL}$ or $U_{\mu i}^{S_LS_L}\equiv U_{\tau
i}^{LL}$ up to a phase, these points correspond to direct flavour
violation.  In this case $NM_{rec}$ is almost at a minimum.
This becomes obvious when we consider the mixing of the
non-standard basis within the standard by performing the rotation
$(\nu^{LL}_\alpha) =
U^{LL\dagger} U^{S_LS_L}(\nu^{S_LS_L}_{\alpha})=W (\nu^{S_LS_L}_{\alpha})$; now
\bea
\left(\begin{array}{c} \nu^{LL}_\mu \\ \nu^{LL}_\tau \end{array}\right)=
\left(\begin{array}{cc} c_\theta c_\phi +s_\theta s_\phi & 
c_\theta s_\phi - s_\theta c_\phi \\ s_\theta c_\phi -c_\theta s_\phi &
s_\theta s_\phi + c_\theta c_\phi \end{array}\right)\left(\begin{array}{c} \nu^{S_LS_L}_\mu \\ \nu^{S_LS_L}_\tau \end{array}\right).\label{eq:mixingeg}
\eea 
For the case of $\sin (2\theta)=1$ the standard basis coincides with
the non-standard when $\cos (\phi)-\sin (\phi)=0$ up to a phase and is directly
flavour violating when  $\cos (\phi)+\sin (\phi)=0$.

\begin{center}
\begin{figure}[p]
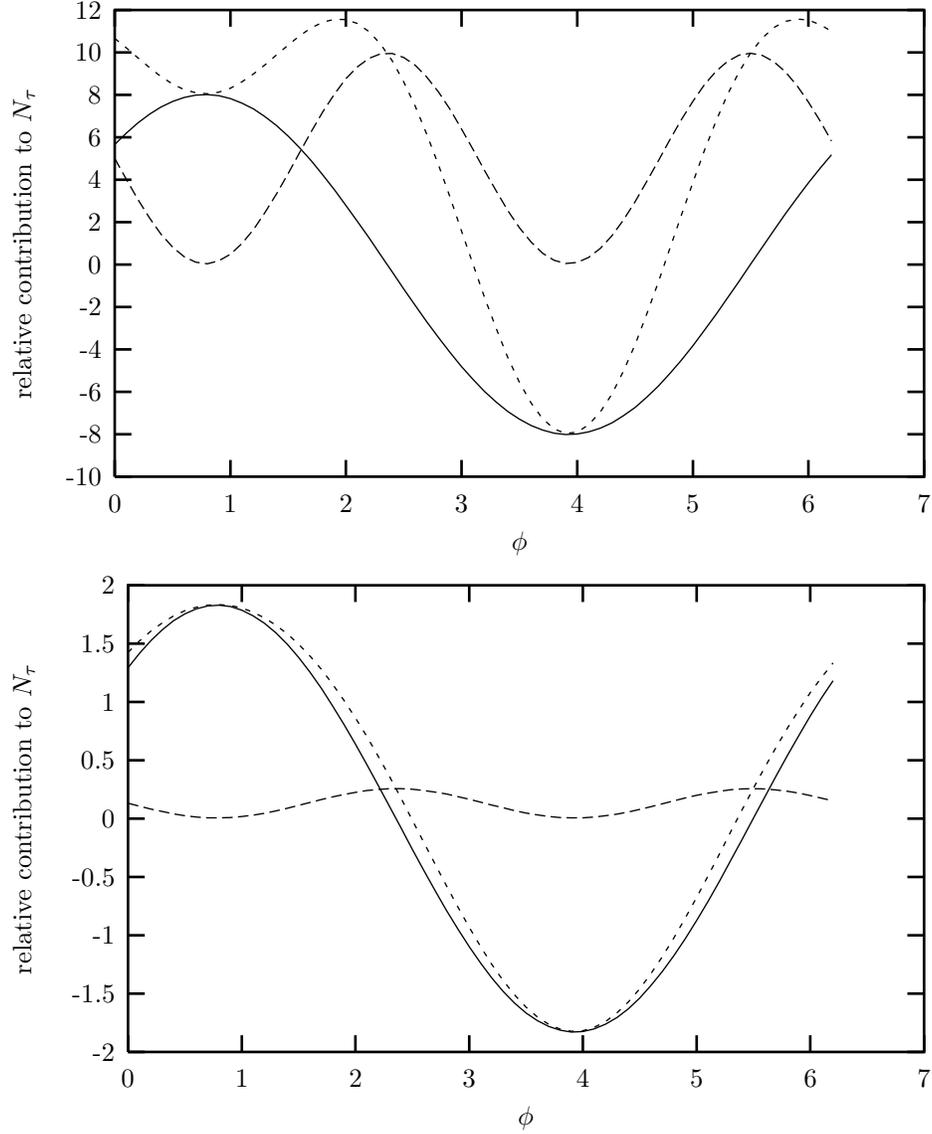

\centering
\input{ntau.SS+SL.E50.ps}
\input{ntau.SS+SL.E20.ps}
\caption{Top panel: Relative contribution to the charged current
event rate for 
the pure scalar term, long dash, the interference term, solid, and the
sum of the two, short dash. With $G^{S_LS_L}=0.01G^{LL}$, $\delta
m^2 = 0.0025 eV^2/c^4$, $L = 732km$
and $E_\mu =50\ {\rm GeV}$.
Bottom panel: Relative contribution to the charged current event rate for
the pure scalar term, long dash, the interference term, solid, and the
sum of the two, short dash. With $G^{S_LS_L}=0.01G^{LL}$, $\delta
m^2 = 0.0025 eV^2/c^4$, $L = 732km$
and $E_\mu =20\ {\rm GeV}$.}
\label{fig:ntau.SS+SL.E20.ps}
\end{figure}
\end{center}

The
two plots also exhibit a strong energy dependence with a much richer
structure evident for the higher beam energy.  This effect may be
understood through a careful examination of the $(\nu_\tau N)$ cross
section.  The interference term is suppressed by a factor of
$E_q^{-1}$ relative to the other terms,  thus its importance is
enhanced at low energies.  In Fig.~\ref{fig:ntau.SS+SL.E20.ps} the
relative contributions to the event rate from the pure scalar and the
interference terms are shown.  The first plot is with a beam energy of
$E_\mu = 50\ {\rm GeV}$ and the second with $E_\mu =20\  {\rm GeV}$.
At high energies the magnitude of the contribution from the pure
scalar term is approximately equivalent to that of the interference
term, this is despite the fact that it is suppressed by a relative
factor of $G^{S_LS_L}$.  At the lower beam energy the interference
term is not yet washed out and dominates over the pure scalar term. 

This simplified analysis indicates that for some values of
$\phi$ a detector mass of $\sim 1000 {\rm kt}$ would be in a good
position to either detect NSI or increase the upper bound on the
non-standard coupling strength.    This result is
to be expected, since at this energy and over a medium base line the
neutrino beam will comprise mostly of non-oscillated $\nu_\mu$'s.  Any
flavour violating non-standard coupling will in essence be picked out over
the standard term.  This example effectively demonstrates the
convenience of this formalism, it allows one to obtain
results for flavour violating couplings, as in Ref.~\cite{Datta:2000ci}, or
flavour diagonal couplings simply by dialling up the appropriate value
of $\phi $.  It also serves as a useful reminder that absent a
specific SM extension the phase of a new interaction can play as
important a role as the coupling strength in oscillation experiments,
 while it is unimportant in spectral tests of NSI.   

\section{Matter enhanced oscillations---two neutrinos}\label{sec:2numat}
We now move on to examine the impact of NSI on the experimental
signature of the matter enhanced $\nu_e\to \nu_\mu$ oscillation
channel.  Again this study is conducted with a generic neutrino
factory in mind. For this calculation only new
interactions during the propagation stage of the oscillation process
are considered,
no new interaction are present at the detector or the source. In
particular a left-chiral scalar interaction coupled with a strength of
$g^{S_LS_L}=0.01 g^{LL}$ is examined.  Since there is no new physics
at the detector or the source it is possible to factor out an `oscillation
probability' in \Eq\ref{eq:idealeventrate}. We will comment further
on this issue later.  

The propagation potential for a neutrino with the SM and left-chiral
scalar interaction present is given by \Eq(\ref{eq:propLagrangian})
and \Eq(\ref{eq:leftpot}) in the mass basis as
\bea
(V^{LL})^\mu\cdot\gamma_\mu(1-\gamma_5) &=& \sqrt{2}G_F U^{LL}_{ei}U^{LL}_{ej}n_e
-\frac{g^{S_LS_L}}{g^{LL}\sqrt{2}}G_F U^{S_LS_L}_{ei}U^{S_LS_L}_{ej} n_e\label{eq:basisstd}
\eea
or in the SM interaction basis
\bea
(V^{LL})^\mu\cdot\gamma_\mu(1-\gamma_5) &=& \sqrt{2}G_Fn_e
-\sum_{\alpha\beta}\frac{g^{S_LS_L}}{g^{LL}\sqrt{2}}G_F
W_{\alpha\tilde{e}}W_{\beta\tilde{e}}  n_e \label{eq:basischange}
\eea
where $W= U^{LL\dagger}U^{S_LS_L}$ was introduced in Eq.~\ref{eq:mixingeg}. We have denoted the non-standard basis with 
the tilde in \Eq(\ref{eq:basischange}).  The propagation potential is
rotated into the SM basis for ease of comparison with the standard
treatment of matter enhanced oscillations,  in this spirit the matrix $W$ is
parameterised in an analogous way to the non-standard vacuum mixing
matrix:
\bea
W&=& \left(\begin{array}{cc}\cos\psi & \sin\psi \\ -\sin\psi &\cos\psi
\end{array}\right)\ .
\eea 
With these considerations the Hamiltonian for neutrinos in
matter is
\bea
\tilde{H} &=& \frac{1}{2E_q}\Biggl[ U^{LL}\left(\begin{array}{cc}
m_1^2 & 0 \\ 0 & m_2^2\end{array}\right)U^{LL\dagger} -
W\left(\begin{array}{cc} \frac{A^{'}}{2}  & 0 \\ 0 & 0\end{array}\right)W^\dagger
+ \left(\begin{array}{cc} A  & 0 \\ 0 & 0\end{array}\right)
\Biggr]\label{eq:2nuham}
\eea
where $A^{'} = 2\sqrt{2}\frac{g^{S_LS_L}}{g^{LL}}G_F n_e E_q$ and $A =
2\sqrt{2}G_F n_e E_q$.  The expression $\tilde{H}$ can be rewritten
as
\bea
\tilde{H} &=&\frac{1}{4E_q}\Biggl[\Sigma + A - \frac{A^{'}}{2} 
+\frac{A^{'}}{2} \left(\begin{array}{cc}-\cos(2\psi) & \sin(2\psi)
\\ \sin(2\psi) & \cos(2\psi)\end{array}\right)\nn\\
&+& A \left(\begin{array}{cc} 1  & 0 \\ 0 & -1\end{array}\right) +
\delta m^2\left(\begin{array}{cc} - \cos(2\theta) &\sin(2\theta) \\ 
\sin(2\theta) & \cos(2\theta)\end{array}\right) \Biggr]\label{eq:2nuham2}
\eea
where $\Sigma = m_2^2 + m_1^2$ and as usual $\delta m^2 = m_2^2 -
m_1^2$.  
Now an angle $\chi$ that rotates $\tilde{H}$ into a diagonal
basis such that $H_M = U_M(\chi)\tilde{H}U(\chi)_M^\dagger$ is found
to be
\bea
\tan(2\chi) = \frac{A^{'}\sin(2\psi) + 2\delta
m^2\sin(2\theta)}{2\delta m^2\cos(2\theta)-2A+2A^{'}\cos(2\psi)}\
.\label{eq:chi}
\eea
There are two distinct eigenvalues of  $H_M$
\bea
m_{LK}  
&=&\!\pm\sqrt{\left(A -\frac{A^{'}}{2}\cos(2\psi)-\delta
m^2\cos(2\theta) \right)^2 +
\left(\frac{A^{'}}{2}\sin(2\psi)+ \delta m^2\sin(2\theta)\right)^2 }\nn\\
&+& \Sigma + A -\frac{A^{'}}{2}\ .\label{eq:nsieigen}
\eea
The Eqs.~\ref{eq:chi} and \ref{eq:nsieigen} elucidate why oscillations
are possible even for vanishing or degenerate vacuum masses. A
non-trivial phase develops and the mass degeneracy is lifted due to
the non-standard interaction in the medium, \emph{ie} the effect is
proportional to $A^{'}$.

The validity of performing an analysis of matter enhanced oscillations
with NSI using a two neutrino model may be questioned.  The
approximations that allow 
the two neutrino analysis to be performed in the standard case need
not hold in the presence of non-standard interactions.  In the standard
oscillation scenario the three neutrino model can be described with
two neutrinos for some values of mixings and mass differences.
The standard $3\times 3$ mixing matrix is written as the product of three
rotations  $U=R(\theta_{23})\cdot
R(\theta_{13})\cdot R(\theta_{12})$.  Typically  the mass
difference $\delta m_{21}^2$ is taken to vanish approximately. This
decouples the 
$R(\theta_{12})$ rotation, in addition the $R(\theta_{23})$ operates
in the $23$-subspace and
commutes with the matter induced term, meaning that only
$\theta_{13}$ is enhanced by the induced potential.  For the
$\nu_e\to\nu_\mu$ channel the standard oscillation probability
becomes:
\bea
P_{e\mu}=\sin^2\theta_{23}\sin^2(2\theta_{13}^M)
\sin^2(\frac{\delta m^2_{13}}{4E_q}\V \vec{L}\V )\ ,\label{eq:bargerprob}
\eea
where $\theta_{13}^M$ is the effective mixing angle in matter and $
\delta m^2_{13}$ is also the effective mass difference in matter.  
Use can be made of \Eq\ref{eq:bargerprob} in studying the effects of
NSI with 
the mass difference and effective mixing angle given by
\Eq\ref{eq:nsieigen} and \Eq\ref{eq:chi} if one assumes that the NSI
basis is only non-diagonal in the $13$ and $12$-subsystems of the
standard basis.  Furthermore one has to assume that the non-standard
equivalent of $R(\theta_{12})$ is approximately equal to the identity.  
For the moment
we acknowledge these assumptions as potential pitfalls and perform an
analysis for the 
sake of gaining an intuitive understanding of the system. The three
neutrino scenario will be examined in the next section. 
For this calculation a standard parameterisation of the $(\nu_\mu N)$
deep inelastic 
scattering cross section may be used since we are assuming no new
physics at the detector, that is~\cite{Freund:2000ti}
\bea
\sigma_{\nu_\mu}(E_q) \simeq 0.67\times 10^{-38} E_q\ {\rm
\frac{cm^2}{GeV}}\ .
\eea
Furthermore the electron neutrino energy distribution has the standard
form
\bea
\frac{d\rho_{\nu_e}}{dxd\Omega_q}= \frac{12}{4\pi} x^2(1-x)\ .
\eea
The matter induced potential can be written as
\bea
A = 2\sqrt{2}G_F Y_e\rho E_q = 1.52\times 10^{-4} {\rm eV^2} Y_e\rho({\rm
g/cm^3}) E_q({\rm GeV} )\ ,
\eea
where $Y_e$ is the electron fraction and $\rho$ is the density of
matter along the neutrino trajectory.  Typically $Y_e\sim 0.5$ in the
earth.  The value of $\rho$ varies between $\sim 3-4.5 {\rm g/cm^3}$
depending on the base length and tilt angle of the experiment. 

\begin{center}
\begin{figure}[p]
\centering
\input{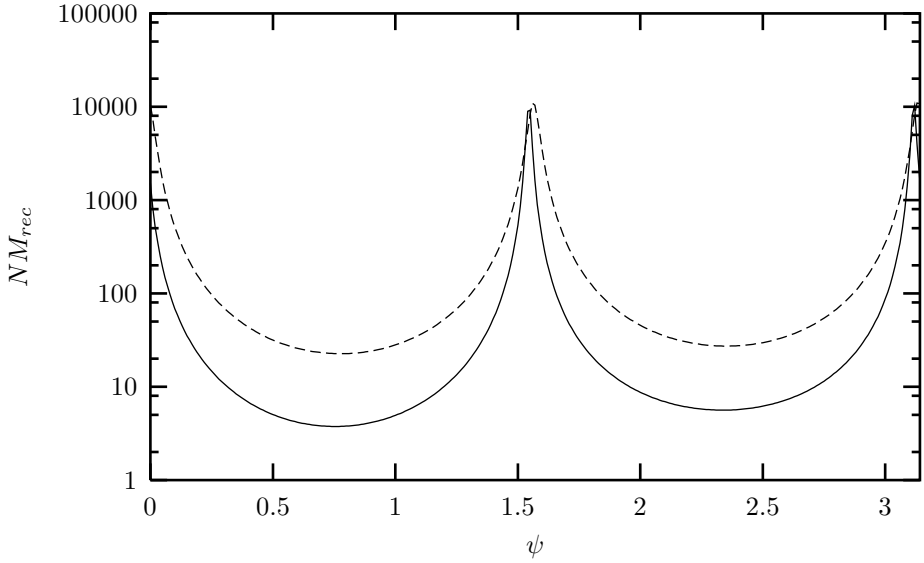}
\caption{The required detector mass-muon number in units of
$2\times 10^{21}\ {\rm kt/yr}$.  The solid line is with $E_\mu =50\
{\rm GeV}$, the
long-dashed line is with $E_\mu=20\ {\rm GeV}$.  The vacuum mass
difference is taken to be positive with $\V\delta m^2\V= 3.5\times
10^{-3}{\rm eV^2/c^4}$, $\V\vec{L}\V=7332{\rm km}$ and
$\sin^2(2\theta)=0.004 $.}
\label{event_rate_matter_phi.ps}
\end{figure}
\end{center}

The dependence on $\psi$ is examined in
Fig.~\ref{event_rate_matter_phi.ps}.  This plot shows two curves one
with $E_\mu = 50\ {\rm GeV}$ and the other with $E_\mu = 20\ {\rm
GeV}$.  The minimum value of $N M_{rec}$ occurs for $\psi=\pi/4$,
as one would expect since this angle corresponds to a maximal mixing
of non-standard flavours in the standard basis.  In fact the behaviour
of the diagonalization angle $\chi$ for high energy neutrinos also
leads one to conclude that the greatest effect will be for $\psi =
\pi/4$.  For large neutrino energy and $\psi=\pi/4$ \Eq\ref{eq:chi}
becomes:
\bea
\tan(2\chi)&\approx&\frac{A^{'}}{-2A}\nn\\
&=&  -0.005\ ,
\eea
for a non-standard coupling strength of $g^{S_LS_L}=0.01g^{LL}$.
Without the NSI in the higher energy limit $\tan(2\chi)\to 0$. These
results are
in contrast to the vacuum oscillations studied in the previous
section.  The case of new
physics at the detector showed the greatest sensitivity when
$\psi=(\frac{\pi}{2},\frac{3\pi}{2})$.  

Given
that the required detector 
mass-muon number varies so dramatically with mixing angle one must
conclude that it is 
not wise to make the approximations that allowed the study of this
system in terms of two neutrinos only.  In addition, note that
the effects of NSI during the propagation stage of the oscillation
process are about three orders of magnitude greater than for a
non-standard interaction of the same strength at either the detection
or production stages.  This implies that one can study NSI during
propagation in isolation of the detector and the source for the matter
enhanced channel, an assumption made without justification in previous
studies~\cite{Fornengo:2001pm, Huber:2001zw, Huber:2002bi}.   

\section{Matter enhanced oscillations---three neutrinos}
The cautionary comment of the previous section stating that the
assumptions required to treat the NSI in a two neutrino framework is
investigated by extending the formalism to allow for three neutrinos.
This is done presently, in addition to quantifying the assertion that
for matter enhanced oscillations NSI at the source and detector may be
neglected. 

Extending the formalism of Section~\ref{sec:2numat} presents no real
hurdles, but an analytic understanding of the results is now
difficult.
  We find it convenient to put the standard and non-standard
interactions with the lepton background on the same footing by
evaluating the propagation potential, Eq.~\ref{eq:basisstd}, in the
mass basis.  The effective mass of a neutrino in matter is found by
evaluating the eigenvalues of the equation of motion in this basis.
The unitary matrix, $\tilde{U}_M$, which diagonalizes the effective
mass matrix are then found algebraically via a method analogous to
that of Ref.~\cite{Cheng:2001mg}.  By making the substitution
$U^{\alpha\beta}\to \tilde{U}^{\alpha\beta}$, where
$\tilde{U}^{\alpha\beta}= U^{\alpha\beta}\tilde{U}_M$, the expression
for the event rate at the detector is equivalent to
Eq.~\ref{eq:totevent}.

  We again investigate the case of one
additional 
non-standard interaction, namely a left-chiral scalar interaction.
The standard and non-standard mixing matrices, $U^{LL}$ and
$U^{S_LS_L}$ respectively,  are parameterised by three different
rotation angles, $\theta_{12}$,  $\theta_{13}$ and  $\theta_{23}$ for
$U^{LL}$ and   $\phi_{12}$,  $\phi_{13}$ and  $\phi_{23}$ for
$U^{S_LS_L}$.  The values of the standard mixing angles are taken to
be
\bea
\sin\theta_{12} &=& 0.53\ ,\\
\sin\theta_{13} & = & 0.03\ ,\\
\sin\theta_{23} &=& 0.71\ ,
\eea
from Ref.~\cite{Freund:2000ti}.  
The numerical evaluation of $NM_{rec}$ proceeds in the same manner as
in Sections~\ref{sec:vacosc}~\&~\ref{sec:2numat}.  When considering
muon production
at the detector as a result of NSI we make the approximation that
$\sigma^{S_LS_L}\sim \sigma^{S_LS_L}$.  For the $\nu_e\to\nu_\mu$
channel the interference term in the cross section is ignored due to
the comparatively small muon mass.

\begin{center}
\begin{figure}[p]
\centering
\input{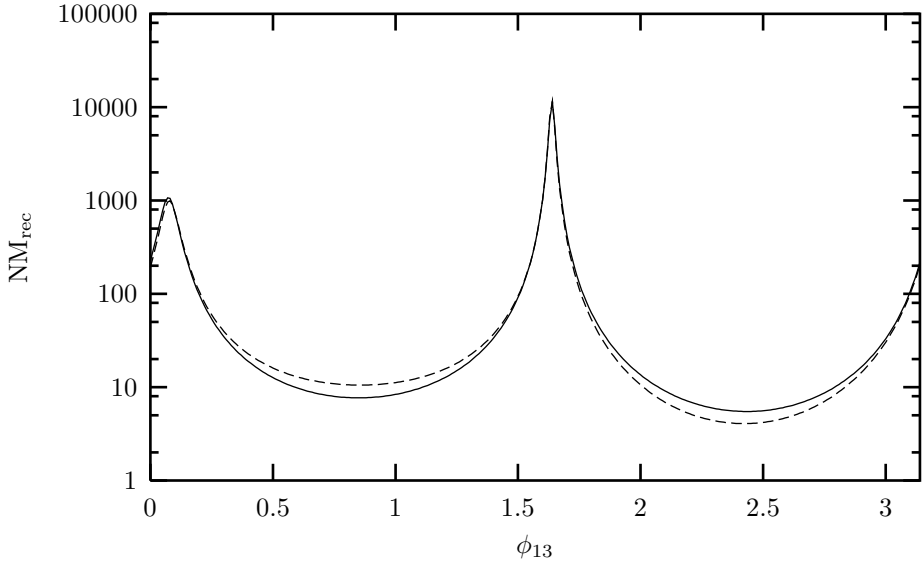}
\caption{The required detector mass-muon number in units of
$2\times 10^{21}\ {\rm kt/yr}$ and for$E_\mu =50\ {\rm GeV}$,  $L=7332\
{\rm km}$, $\phi_{12}=0$, $\phi_{23}=\theta_{23}$ and
$g^{S_LS_L}=0.01g^{LL}$.  The 
solid line is the event rate with no new physics at the detector.
While the dashed line is the event rate with a non-standard coupling of
$G^{S_LS_L}=0.01G^{LL}$ at the detector. 
}
\label{3nu.ps}
\end{figure}
\end{center}

\begin{center}
\begin{figure}[p]
\centering
\input{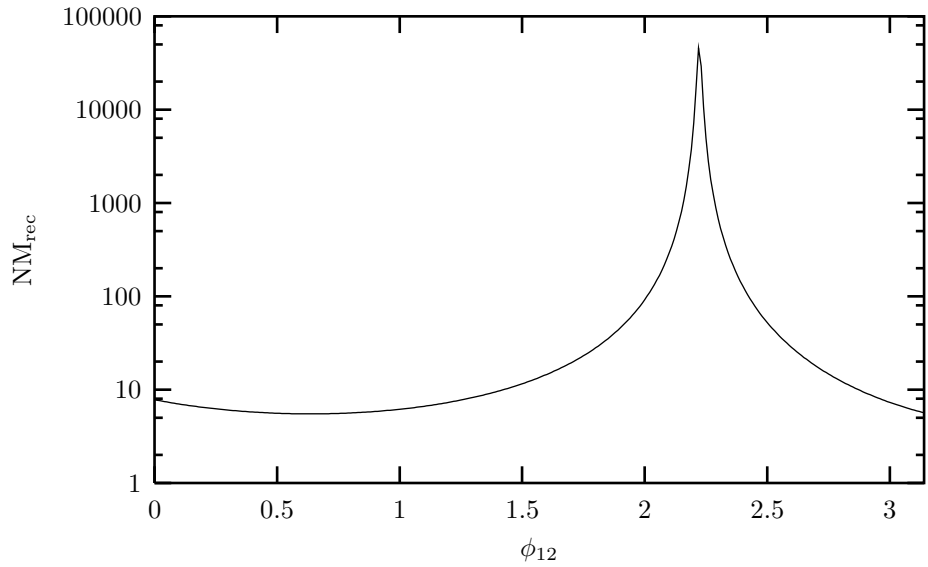}
\caption{The required detector mass-muon number in units of
$2\times 10^{21}\ {\rm kt/yr}$ as a function of mixing angle
$\phi_{12}$. 
The other parameters are $E_\mu =50\ {\rm GeV}$,  $L=7332\
{\rm km}$, $\phi_{13}=\pi/4$, $\phi_{23}=\theta_{23}$ and
$g^{S_LS_L}=0.01g^{LL}$. In this plot 
we have assumed no new 
physics at the detector. 
}
\label{phi13.ps}
\end{figure}
\end{center}

In Figure~\ref{3nu.ps} \& \ref{phi13.ps} we have plotted the variation of ${\rm
NM_{rec}}$ as a function of $\phi_{12}$ and $\phi_{13}$ respectively.
Note that since we are only considering the effects of the electron
background and, for the time being, no new physics at the detector the
angle $\phi_{23}$ decouples from the calculation as per the discussion
of the previous section.  As was expected the variation of ${ NM_{rec}}$ in
Figure~\ref{3nu.ps} takes much the same form as in
Figure~\ref{event_rate_matter_phi.ps} when $\phi_{12}=0$, or
$R(\phi_{12})$ is diagonal.  However when we choose $\phi_{13}=\pi/4$
and vary $\phi_{12}$ the results are dramatic.  While not
unexpected these results  serve to highlight the point that any experimental
analysis aiming to place bounds on the couplings of NSI must consider
the full three neutrino system.  This applies even for cases where the
standard treatment may be done in a two neutrino framework.

Also shown in Figure~\ref{3nu.ps} are the effects of including
non-standard physics at the detector in addition to the propagation
stage.  We see that the assertion made that the results of new physics at the
detector are negligible when compared with effects of new physics
during propagation are in this case justified.
\section{Concluding remarks}
In this paper we have united parameterisations of physics beyond the
SM such as those of the Michel parameters and
non-standard $\beta$-decay couplings of precision weak physics experiments with
the phenomenon of neutrino oscillations.  As such a  framework
for analysing specific theories, such as Supersymmetry, leptoquarks
etc.~has been established. This was achieved by developing a
scattering theory of the production, 
propagation and detection processes.  In doing so we challenged the
notion of what is termed  a  
neutrino flavour state.  This was necessitated by the fact that neutrino
flavour states do not have a definite mass as is required of
intermediate states in a scattering theory, and phenomenologically
there is no compelling reason to assume that new interactions couple
to the same linear combination of mass states as the V-A interaction. 
As a result of this formalism we were able to
show explicitly the known results that flavour violation or
partial flavour 
violation can not lead to oscillation phenomena in a vacuum. And, in
contradistinction,  that
flavour violation or partial flavour violation in matter can lead to
oscillations 
even for degenerate vacuum masses.  The condition for the latter
result is that the non-standard interaction be non-diagonal in the
standard basis. 
In addition we have shown that in the ultra-relativistic limit, production
and detection processes with opposite chirality decouple, that is
right-handed neutrinos become fully sterile from their left-handed
counterparts.   However in matter a
right-handed potential acts as an effective mass linking the two
states.   

The field theoretic description of neutrino oscillations was used to
perform some simple calculations for a generic neutrino factory.
This type of experiment was chosen as a convenient starting point due the 
high intensity and energy of the beam with which the
experiments are to be conducted.  The beam energy is an important
consideration since the field theory was derived in the limit that
$\delta m^2\V\vec{L}\V /2E_q<1$.   

For the vacuum oscillation case we examined the $\nu_\mu\to\nu_\tau$
channel with a left chiral scalar coupling of strength
$G^{S_LS_L}=0.01G^{LL}$.  By allowing the non-standard mixing angle, $\phi$, to
vary we were able to observe  flavour conserving and all flavour violating
scenarios.  We found the greatest effect, at high energies, was for
flavour violating interactions which effectively picked out the
non-standard term 
in the cross section.  Just above the $\tau$-production threshold
flavour conserving interactions play a more important role.  

The matter enhanced $\nu_e\to\nu_\mu$ channel was also investigated.
Two important differences from the vacuum case were observed.
Although the coupling strength of the NSI was the same as that for the
one used in the vacuum oscillation study the experimental sensitivity
was up to three orders of magnitude greater, indicating that it is
safe to ignore new physics at the detector and source for the matter
enhanced channels.  We also note that the greatest effect when varying
the non-standard mixing angle,
$\psi$, did not correspond to direct flavour violation, rather when the
non-standard interaction was maximally mixed in the standard basis.
In the conventional parlance this would correspond to a decay to a
superposition of flavour states.

The focus of this paper has been one of developing a solid
framework for future work.  Future calculations will concentrate on more
realistic experimental simulations than the ones presented here and on
performing an analysis of the impact of a right-handed interaction on
neutrino propagation through matter.
\bibliography{Bibleo}

\pagebreak

\end{document}